\documentclass[12pt]{article}
\usepackage[dvips]{graphicx}

\newcommand{\be}{\begin{equation}}
\newcommand{\ee}{\end{equation}}
\newcommand{\bea}{\begin{eqnarray}}
\newcommand{\eea}{\end{eqnarray}}

\begin{document}
\begin{titlepage}
 
 
\vspace{1in}
 
\begin{center}
\Large
{\bf Cyclical Behaviour in Early Universe Cosmologies}
 
\vspace{1in}

\normalsize

\large{Andrew P. Billyard$^{1}$, Alan A. Coley$^{1,2}$
 \& James E. Lidsey$^{3}$}

\normalsize
\vspace{.7in}

$^1${\em Department of Physics, \\ Dalhousie University, Halifax, NS, B3H 3J5,
Canada} \\

\vspace{.1in}

$^2${\em Department of Mathematics and Statistics, \\ Dalhousie University, Halifax, NS, B3H 3J5,
Canada} \\

\vspace{.1in}

$^3${\em Astronomy Unit, School of Mathematical Sciences, \\ Queen Mary \& Westfield, 
Mile End Road, London, E1 4NS, UK} \\

\vspace{.1in}

\end{center}
 
\vspace{.5in}
 
\baselineskip=24pt
\begin{abstract}
\noindent

We study early universe cosmologies derived from a scalar--tensor 
action containing cosmological constant terms and massless fields.
The governing equations can
be written as a dynamical system which contains no past or future
asymptotic equilibrium states (i.e. no sources
nor sinks). This leads to dynamics with very interesting mathematical behaviour
such as 
the existence of heteroclinic cycles. The corresponding cosmologies
have novel characteristics, including cyclical and bouncing behaviour 
possibly indicating chaos. We discuss the connection between these
early universe cosmologies and those derived
from the low--energy string effective action.

\end{abstract}

\end{titlepage}

\ 

\section{Introduction}

In this paper we consider the qualitative 
dynamics of a class of spatially flat, 
scalar--tensor cosmological models
derived from the action
\begin{equation}
\label{generalaction}
S=\int d^4 x \sqrt{-g} \left\{ e^{-\Phi} \left[ R +\left( \nabla \Phi
\right)^2 -\frac{1}{2} e^{2\Phi}
\left( \nabla \sigma \right)^2 -2\Lambda\right] - \Lambda_{\rm M} \right\}
\end{equation}
where $R$ is the Ricci curvature scalar of the space--time with metric 
$g_{\mu\nu}$, $g \equiv {\rm det}g_{\mu\nu}$, $\{ \Lambda ,
\Lambda_{\rm M} \}$ are constants and $\{ \Phi
 , \sigma \}$ represent scalar fields. 
The dynamics of these cosmological models has very interesting
mathematical properties. In particular, there are no 
asymptotically attracting equilibrium states in the 
phase space and this may lead to important physical consequences.

The form of 
action (\ref{generalaction}) can be partially motivated from string theory,
which is the most promising 
candidate for a unified theory of the fundamental 
interactions \cite{gsw,polchinski}. When $\Lambda_{\rm M}$ vanishes, 
Eq. (\ref{generalaction}) represents the truncated effective 
four--dimensional action of the 
Neveu--Schwarz/Neveu--Schwarz (NS--NS) sector of the theory \cite{gsw}.
The scalar field, $\Phi$, represents the dilaton field and 
the axion field, $\sigma$, is the Poincar\'e dual of the 
antisymmetric two--form potential. The constant, $\Lambda$, may 
be interpreted in terms of the central charge deficit 
of the string theory and can be negative. 
The evolution of the very early
universe immediately below the string scale may 
have been determined by an effective action of this form. 
The dynamics of the spatially flat 
and homogeneous cosmologies in the case $\Lambda_{\rm M}=0$
was presented in Ref. \cite{bclQ}.
One of the main 
purposes of the present work is to determine the effects of introducing 
a non--trivial cosmological constant, $\Lambda_{\rm M}$, 
that does not couple directly to the dilaton field. Such a term 
represents a vacuum 
energy contribution to the energy--momentum tensor.
A discussion of the spatially flat 
and homogeneous cosmologies in the case $\Lambda=0$
was presented in Ref. \cite{bcl}, and a
partial analysis of the dynamics with a non-vanishing axion field when
both $\Lambda$ and $\Lambda_{\rm M}$ are non-zero 
was investigated in \cite{billyard}.

\section{Analysis}

We assume that the metric corresponds to the spatially 
flat, Friedmann--Robertson--Walker (FRW) universe: 
$ds^2 =-dt^2 +e^{2\alpha (t)}dx_idx^i$. 
Substituting this {\em ansatz} into the 
action (\ref{generalaction}) and integrating over the spatial variables 
then yields the reduced action
\begin{equation}
\label{taction}
S=\int dt ~~e^{3\alpha} \left\{ e^{-\Phi} \left[ 6\dot{\alpha}\dot{\Phi} 
-6\dot{\alpha}^2 -\dot{\Phi}^2 +
\frac{1}{2} e^{2\Phi} \dot{\sigma}^2 -2\Lambda\right] -\Lambda_{\rm M} \right\}   ,
\end{equation}
where a dot denotes differentiation 
with respect to cosmic time, $t$. 
The Friedmann constraint derived from Eq. (\ref{taction}) is given by  
\begin{equation}
\label{frr}
3\dot{\alpha}^2 -\dot{\varphi}^2 +2\Lambda
+\frac{1}{2} \dot{\sigma}^2 e^{2\varphi +6 \alpha} + 
\Lambda_{\rm M}  e^{\varphi +3\alpha} =0,   
\end{equation}
in terms of the  `shifted' dilaton field,  $\varphi \equiv \Phi - 3\alpha$.

A generalization to the spatially flat, Bianchi type I cosmology 
may also be considered. This effectively results in the introduction of 
two massless scalar fields into the reduced action 
(\ref{taction}). These fields parametrize the shear of the models.
Similar degrees of freedom also arise 
when considering the 
toroidal compactification of higher--dimensional theories. 
Although we do not consider these extra fields in this paper, 
their overall contribution to the dynamics can be modelled by 
introducing a single modulus 
field, $\dot\beta^2 \equiv \sum_i \dot\beta^2_i$, into
the reduced action (\ref{taction}) \cite{bcl}, and their inclusion 
could be important in the discussion of chaotic behaviour.

\

\noindent
{\em Zero central charge deficit}

\

We first consider the case $\Lambda=0$ \cite{bcl}. 
We assume that
$\Lambda_{\rm M} > 0$, and employ the generalized Friedmann
constraint equation (\ref{frr}) to eliminate the axion field from 
the system.  The resulting 
field equations may then be simplified by introducing the new
variables and time coordinate\footnote{We assume that $\psi>0$; 
the case 
$\psi<0$ is related to a time-reversal of the system and the qualitative 
mathematical behaviour is similar (although the physical interpretation is
quite different).}:
\begin{equation}
x \equiv \frac{\sqrt{3} \alpha'}{\psi} , \quad
z \equiv \frac{\Lambda_{\rm M}}{\psi^2}, \quad
\frac{d}{d\Theta} \equiv \frac{1}{\psi} \frac{d}{d\theta} \equiv \frac{1}{\psi} e^{-(\varphi +3\alpha )/2} \frac{d}{dt} ,
\end{equation}
where a prime denotes differentiation with 
respect to $\theta$ and $\psi \equiv  \varphi'$.
The Friedmann constraint yields 
\begin{equation}
1-x^2-z\geq0,
\end{equation} 
from which it follows  
that the phase space is
bounded with 
\begin{equation}
0\leq \{ x^2,z\}\leq 1.
\end{equation}
The
invariant set $1-x^2-z=0$ corresponds to a trivial axion field. 


The cosmological field equations for the isotropic FRW model 
can now be expressed in terms of the plane system:
\begin{eqnarray}
\label{dxdTh}
\frac{dx}{d\Theta} &= & (x+\sqrt{3})[1-x^2-z] +\frac{1}{2} z[x-\sqrt{3}] \\
\label{dzdTh}
\frac{dz}{d\Theta} &= & 2z\left\{[1-x^2-z] 
-\frac{1}{2}(1-z-\sqrt{3}x)\right\} .
\end{eqnarray}
The equilibrium points of this 
system are $L^+_{(-)}$ $(x,z=-1,0)$, $L^+_{(+)}$ $(1,0)$ and
$S^+$ $(-1/3\sqrt{3},16/27)$. The first two are 
saddles and $S^+$ is a repelling focus. The functional form of these 
solutions was presented and discussed in Ref. \cite{bcl} 
and 
the phase portrait is given in Fig. 1. We note that the exact solutions
corresponding to all of the equilibrium points are self-similar
cosmological models \cite{WE}.
 \begin{figure}[htp]
  \centering
   \includegraphics*[bb=182 235 512 536, width=3in]{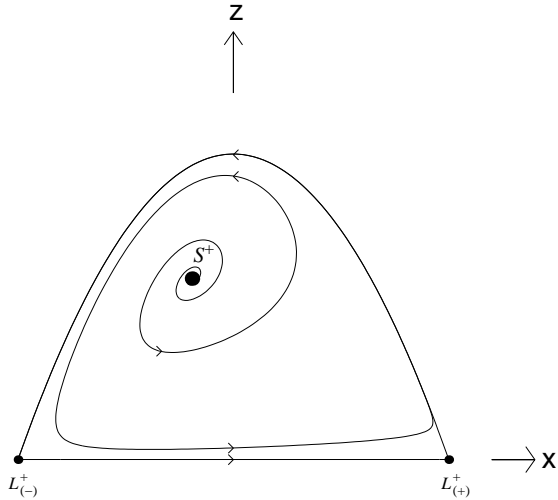}
   \caption{{\em Phase portrait of the system
(\ref{dxdTh})--(\ref{dzdTh}), corresponding to the isotropic 
FRW model with $\Lambda_{\rm M} >0$ and $\Lambda=0$.
We shall adopt the convention that large black dots
represent sources (i.e., repellers), large grey-filled dots represent
sinks (i.e., attractors), and small black dots represent saddles.  
Note that in
this phase space orbits are future asymptotic to a heteroclinic cycle.}}
 \end{figure}

We see from Fig. 1  that the 
orbits are future asymptotic to a {\em heteroclinic cycle}. 
This is comprised 
of the two saddle equilibrium points $L^+_{(-)}$ and $L^+_{(+)}$ and the single
(boundary) orbits in the invariant sets $z=0$ $(\Lambda_{\rm M}=0)$
and $1-x^2-z=0$ $(\dot{\sigma}=0)$ joining
$L^+_{(-)}$ and $L^+_{(+)}$. 
Hence,  the orbits
exhibit {\em cyclical} behaviour \cite{bcl}. 
For each `cycle', an orbit is quasi--stationary in the neighbourhood 
of the saddle point
$L^+_{(-)}$. It then  shadows the orbit in the invariant set $z=0$ 
as it moves rapidly towards the equilibrium point 
$L^+_{(+)}$. It settles into another quasi--stationary phase
close to $L^+_{(+)}$ and eventually moves quickly  
back to $L^+_{(-)}$ shadowing the orbit in the invariant set $1-x^2-z=0$.
It is important to emphasize that the orbits move progressively 
closer towards the two saddles, $L^+_{(\pm)}$, after  
the completion of each cycle. Thus, the motion is {\em not} 
periodic and a given orbit spends 
more and more time in the
neighbourhood of these equilibrium points. 

The physics behind the
cyclical nature of these orbits is as follows. 
The sign of the variable $x$ determines whether 
the universe is expanding or contracting. The value of this 
variable passes through zero 
during each cycle. This behaviour arises because 
the  cosmological constant effectively
resists the expansion of the universe, but the axion field 
has the opposite effect. Since the energy 
density of the latter scales as $\dot{\sigma}^2 \propto e^{-6\alpha}$,
it is negligible when the spatial volume of the universe 
is large. Consequently, the cosmological constant 
forces the expanding universe to recollapse. However, 
the axion field inevitably becomes dominant and reverses this collapse, 
causing the 
universe to enter into a new expanding phase. The 
process is then repeated 
and the interplay between the two opposing trends results
in a universe that undergoes a series of 
bounces. 


\

\noindent
{\em Non-zero central charge deficit}

\

We now  consider the case $\Lambda \ne 0$.
We again employ Eq. (\ref{frr}) to eliminate the $\dot{\sigma}^2$ term
from the field equations, and make the following definitions:
\begin{equation}
x\equiv \frac{\sqrt3\dot\alpha}{\xi}, \quad
y\equiv \frac{ -2\Lambda}{\xi^2}, \quad
z\equiv \frac{\Lambda_{\rm M}  e^{\varphi +3\alpha}}{\xi^2}, \quad
u\equiv \frac{\dot\varphi}{\xi}, \quad
\frac{d}{dt}\equiv \xi \frac{d}{dT}.\label{TheDefs}
\end{equation}
We assume that $\Lambda<0$ ($\Lambda_{\rm M}>0$) and we 
define $\xi^2 = \dot\varphi^2-2\Lambda$.
The generalized
Friedmann constraint equation (\ref{frr}) now yields
$0 \leq x^2 + z \leq 1$, so that all
variables are bounded: $0\leq \{x^2,y,z,u^2\}\leq 1$.
{}From the definition of $\xi$, $y$ is given by $u^2+y=1$. The resulting
three-dimensional system therefore becomes:
\begin{eqnarray}
\label{dx}
\frac{dx}{dT} & = & \sqrt 3 \left( 1-x^2-\frac{3}{2}z  \right) 
                     + xu \left( 1-x^2-\frac{1}{2}z \right), \\
\label{du}
\frac{du}{dT} &=& (1-u^2) \left(x^2+\frac{1}{2}z\right) > 0,\\
\label{dz}
\frac{dz}{dT} &=& z \left[u\left( 1-2x^2-z \right)+\sqrt3 x\right].
\end{eqnarray}

The invariant sets $x^2+z=1$, $z=0$, $u^2=1$ define the boundary of
the phase space and it is important to note that 
the variable $u$ is {\em monotonically
increasing}. This ensures that there are no 
closed or recurrent orbits in the phase space. The equilibrium
points of the system are all saddles:  
$S^\pm$ $(x,u,z=\mp1/\sqrt{27},\pm 1,16/27)$,  
$L^+_{(\pm)}$ $(\pm1,1,0)$ and $L^-_{(\pm)}$ $(\pm1,-1,0)$. 
The points $L^+_{(\pm )}$ represent power--law cosmologies 
with $\dot{\varphi} >0$, where 
only the dilaton field is non--trivial, i.e., 
the axion field and cosmological constant terms are 
dynamically negligible. 
These solutions are termed `dilaton--vacuum' solutions
and have an analytical form given by $e^{\alpha} \propto 
t^{\pm 1/\sqrt{3}}$ and $e^{\Phi} \propto t^{-1 \pm \sqrt{3}}$.
The points $L^-_{(\pm )}$ are 
the corresponding solutions where $\dot{\varphi}<0$. 
The phase portrait is given in Fig. 2. 
 \begin{figure}[htp]
  \centering
   \includegraphics*[width=5in]{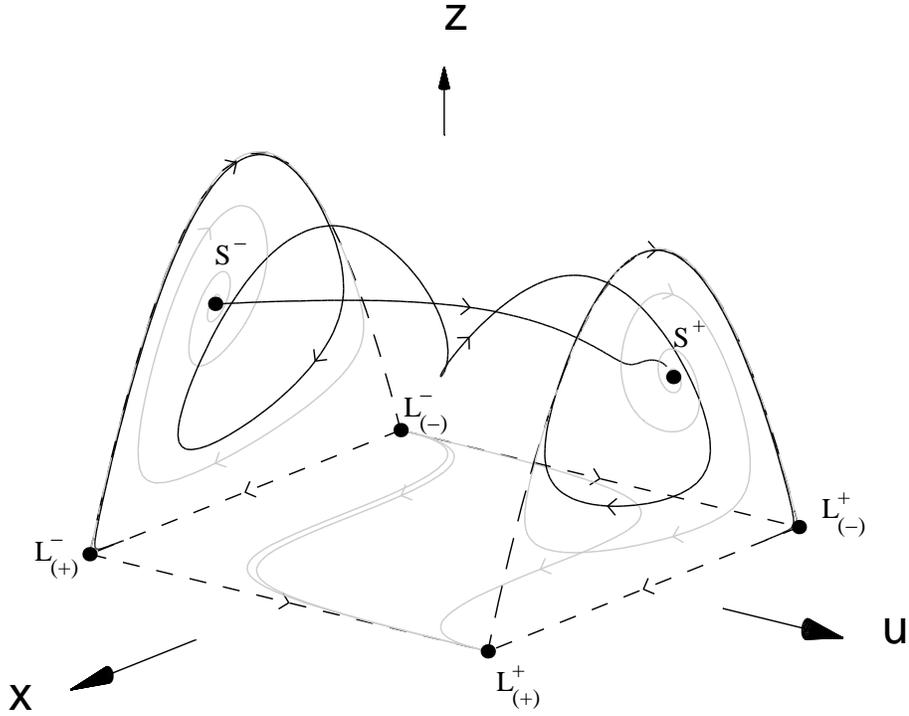}
  \caption{{\em Phase diagram of the system
(\ref{dx})-(\ref{dz}) for $\Lambda<0$ and  $\Lambda_{\rm M}>0$, 
where $\dot\varphi>0$ is
assumed.  See caption to Figure 1. Grey lines represent typical
trajectories found within the two-dimensional invariant sets, dashed
black lines are those trajectories along the intersection of the
invariant sets, and solid black lines are typical trajectories within
the full three-dimensional phase space.}}
 \end{figure}

In this case there are {\em no sinks} and {\em no sources} in the full
three--dimensional phase space. 
Since the variable $u$ (and hence $\dot\varphi$)
is monotonically increasing, 
solutions generically asymptote in both
the past and future towards the invariant sets $u=\pm1$. 
These both include an
heteroclinic cycle and this implies 
that generically the solutions exhibit similar asymptotic behaviour at
both early and late times to that discussed 
above. (For example, to the future the orbits in the 
three--dimensional phase space shadow the orbits in the
two--dimensional invariant set $\Lambda=0$.)
The orbits interpolate between the dilaton--vacuum solutions
corresponding to the saddle points $L^-_{(\pm)}$ in the past and the
dilaton--vacuum solutions corresponding to the saddle points
$L^+_{(\pm)}$ in the future. The effect on the dynamics of the 
cosmological constant, $\Lambda_{\rm M}$, is
significant at both early and late times.  The points
$S^\pm$ correspond to the equilibrium point $S^+$ in Fig. 1, 
but unlike in the $\Lambda=0$ case in which  $S^+$ is a repelling
focus, in the three--dimensional phase space they are saddles 
and hence they do not play a
primary role in the asymptotic behaviour.

{}From a physical point of view, the cyclical 
behaviour arises due to the complex interplay 
between the axion field and the cosmological constant terms. 
The universe continues to undergo a succession of 
bounces between expanding and contracting 
phases due to the axion field 
and vacuum energy $\Lambda_{\rm M}$. However, 
the inclusion of the central charge deficit, $\Lambda$, 
causes $\dot{\varphi}$ to ultimately change sign. 
Thus, the asymptotic behaviour in the future is related 
to a time--reversal of the asymptotic behaviour in the past.
We remark that the dilaton--vacuum solution, $e^{\alpha} \propto 
t^{-1/\sqrt{3}}$, corresponding to 
the point $L^+_{(+)}$ is inflationary over the range $t<0$, 
because the expansion is accelerating. In this case, the accelaration 
is driven by the kinetic energy of the dilaton field.
(For a recent review of the cosmological 
significance of these solutions see, e.g., Ref. \cite{review}).

Finally, we make some brief remarks on the case $\Lambda>0$. 
We can define $\xi \equiv \dot\varphi$ and 
consider the subset $\dot\varphi\geq0$.  Introducing 
normalized variables as before yields a
three--dimensional, compact system of autonomous, ordinary
differential equations. We have completed  
a full dynamical analysis of this system, but we only
describe the main features here.
There is a non-hyperbolic equilibrium point, $C^+$, which
can be shown to be a (global) source, since 
$y$ is a monotonically
decreasing function. This point 
represents a static universe, where the dilaton field is evolving 
linearly with time and the axion field and $\Lambda_{\rm M}$ are 
dynamically negligible. 
There are also two saddle points, $L^+_{(\pm)}$, 
which represent
dilaton--vacuum solutions; these are analogues 
of the saddles that appear above.
Again, there is also a saddle $S^+$. We stress that
there are no sinks in the phase space.
Therefore, trajectories generically
asymptote into the past towards $C^+$, 
and then spiral away towards the heteroclinic cycle
in the
invariant set $y=0$ containing the saddle points
$L^+_{(-)}$ and $L^+_{(+)}$. The phase space is depicted in Fig. 3.
 \begin{figure}[htp]
  \centering
   \includegraphics*[width=3.5in]{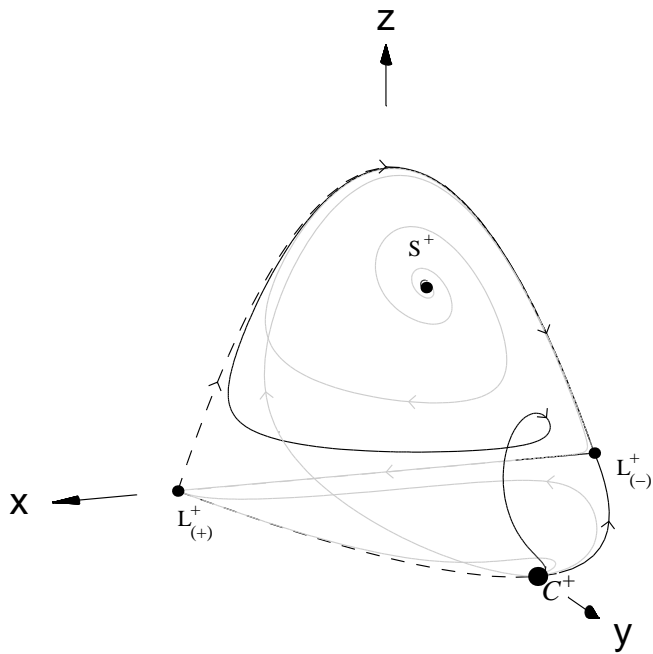}
  \caption{{\em Phase diagram of the system
with $\Lambda>0$ and $\Lambda_{\rm M}>0$,  
where $\dot\varphi>0$ is
assumed. See captions to Figures 1 and 2.}}
 \end{figure}

\section{Discussion}

The most important mathematical feature of the models we have
considered is the {\em cyclical} behaviour 
that arises due to the existence of a
heteroclinic cycle. This is of great physical significance, 
because it might be an  indicator of possible chaotic
behaviour. The solutions
interpolate between different 'Kasner-like'
dilaton--vacuum,  power--law models,
undertaking cycles between the saddles in the three--dimensional phase
space. This is similar to the dynamical behaviour that occurs in spatially homogeneous
Bianchi cosmological models \cite{WE,hobill}. 
The question of chaos in anisotropic Bianchi type IX string 
cosmologies has been considered \cite{chaos}. It 
was shown that since the axion and
dilaton fields behave collectively as 
a stiff perfect fluid, the system oscillates only a
finite number of times. Consequently, there is no Mixmaster-type chaos in these
models \cite{hobill,LeB}.  
This is to be expected since it 
is known 
that the admission of stiff fluid matter causes chaos to cease
\cite{b9stiff}.
In contrast to the 
anisotropic Bianchi type IX cosmologies, 
however, 
the models described above contain cosmological
terms, e.g., an effective dilaton potential and a dynamically important
axion field. It is these intrinsically stringy effects that give
rise to chaotic behaviour and 
this chaos has a different origin to the chaotic behaviour that arises in 
general relativistic models. 
On the other hand, there may be some connection with models 
that contain Yang--Mills fields. 
It is known that chaotic oscillations occur for such fields
\cite{BL} and, moreover, it was shown 
in \cite{b9stiff} that the oscillations that are suppressed by a single 
massless scalar field can be restored 
by coupling an electromagnetic
field to a Brans-Dicke type field. This model is related 
to a scalar field model with an exponential potential \cite{exp}
and, consequently, is also 
related to string theory cosmological models \cite{LP}.

There are a number of outstanding issues that need to be addressed
regarding the possible existence of chaos in string cosmology. First, the
chaotic behaviour depends crucially on the dimensionality  of
spacetime and on
the product manifold structure of the extra dimensions \cite{chaos}. 
In particular,
superstring theories are formulated in $D=10$ spacetime dimensions,
while $M$-theory, with its low-energy supergravity limit, is
an eleven--dimensional theory \cite{witten}. 
Second, the low energy effective action is only valid in 
the perturbative regime of weak coupling and small curvature.
In general, it may be necessary  to study chaos within 
the context of a full
non-perturbative formulation of the theory, but at present 
such a formulation is unknown.
Nevertheless, if chaotic behaviour occurs
at the level of the 
effective action, it is to be expected that similar 
behaviour should arise in the non--perturbative
regime. Finally, there is the
question of what will happen if inhomogeneities are introduced. 
Again, such effects will be most unlikely to lead to any suppression of
chaotic behaviour and will perhaps 
make chaos even more predominant \cite{chaos}.

There are other questions which are
important in early  universe cosmology in general, and in
string cosmology in particular. The questions of whether
cosmological models can isotropize and/or inflate (and if they
can inflate whether there is a graceful exit from inflation) are of
great importance \cite{diam}. The
techniques utilized in this paper can easily be adapted to study the possible
isotropization in more general spatially homogeneous but anisoptropic string
cosmological models. Inflationary properties of simple string cosmologies have been discussed above. However, 
a chaotic cosmological regime might either be an alternative to inflation
or, perhaps more importantly, could work in tandem with an inflationary mechanism 
\cite{YST} to produce new
interesting physical phenomena. For example, 
a chaotic regime due to dissipative effects
or chaotic mixing \cite{LMP} could possibly be an alternative to
inflation as a cause of homogenization and isotropization. 
This might alleviate the problems of initial conditions
in inflation. This last point has been addressed 
in \cite{CSS}, 
where it was suggested that there would be sufficient time for a
compact, negatively--curved  universe to homogenize since 
chaotic mixing smooths out primordial fluctuations in a
pre-inflationary period.

\end{document}